\def\preprint{0}                
\def\preprint{1}                
\def\comment#1{} 
\if\preprint1 
        \documentclass[usenatbib]{mn2e} 
        \usepackage{times} 
        \usepackage{graphicx} 
        \usepackage{calc} 
 
\else 
        \documentstyle[astrop-bib,referee,times]{mn} 
        \newcommand{\includegraphics}[1]{} 
\fi

\def\oversim#1#2{\lower0.5pt\vbox{\baselineskip0pt \lineskip-0.5pt 
     \ialign{$\mathsurround0pt #1\hfil##\hfil$\crcr#2\crcr\sim\crcr}}} 
\def\lsim{\mathrel{\mathpalette\oversim<}}    
\def\aap{{\rm A\&A}} 
\def\mnras{{\rm MNRAS}} 

\title[Carbon stars  
in the Sgr dSph]{Metal-rich carbon stars in the Sagittarius Dwarf Spheroidal  
galaxy} 
\author[E. Lagadec et al.]{Eric Lagadec$^{1}$ \thanks{E-mail: 
eric.lagadec@manchester.ac.uk}, 
               Albert~A.~Zijlstra$^1$, 
   G.C. Sloan$^2$,  Peter R. Wood$^3$, Mikako Matsuura$^{4,5}$,\newauthor 
 Jeronimo Bernard-Salas$^2$, J.A.D.L. Blommaert$^6$,  
M.-R. L. Cioni$^7$, M.W Feast$^{8,11}$,\newauthor 
M.A.T. Groenewegen$^9$,  Sacha Hony$^{10}$, 
 J.W. Menzies$^{11}$, J.Th. van Loon$^{12}$,\newauthor  
 P.A. Whitelock$^{8, 11,13}$\\ 
\\ 
$^1$Jodrell Bank Centre for Astrophysics, The University of Manchester,  
School of Physics \&\ Astronomy,  Manchester M13 9PL, UK\\ 
$^2$Department of Astronomy, Cornell University, 108 Space Sciences  
            Building, Ithaca NY 14853-6801, USA\\  
$^3$Research School of Astronomy and Astrophysics,  
          Australian National University,  Cotter Road, Weston Creek, 
          ACT 2611, Australia\\ 
$^4$  Division of Optical and IR Astronomy, National Astronomical Observatory  
     of Japan, Osawa 2-21-1, Mitaka, Tokyo 181-8588, Japan\\ 
$^5$ Department of Physics and Astronomy, University College London,  
Gower Street, London WC1E 6BT, UK\\ 
$^6$Instituut voor Sterrenkunde, K.U.Leuven, Celestijnenlaan 200D,  
3001 Leuven, Belgium\\ 
$^7$Centre for Astrophysics Research, University of Hertfordshire,  
Hatfield AL10 9AB, UK\\ 
$^8$ Astronomy Department, University of Cape Town, 7701, Rondebosch,  
South Africa\\ 
$^9$ Koninklijke Sterrenwacht van Belgie, Ringlaan 3, 1180 Brussels, Belgium\\ 
$^{10}$ CEA, DSM, DAPNIA, Service d'Astrophysique, CEA Saclay,  
91191 Gif-sur-Yvette Cedex, France \\ 
$^{11}$ South African Astronomical Observatory, PO Box 9,  
7935 Observatory, South Africa\\ 
$^{12}$Astrophysics Group, School of Physical \&\ Geographical Sciences,  
Keele University, Staffordshire ST5 5BG, UK\\ 
$^{13}$ National Astrophysics and Space Science Programme,  
 Department of Mathematics and Applied Mathematics\\ 
 University of Cape Town, 7701 Rondebosch, South Africa\\ 
} 
\pubyear{2007} 
 
\begin{document} 
 
\date{Accepted . Received}

\pagerange{\pageref{firstpage}--\pageref{lastpage}} \pubyear{2002} 
 
\maketitle 
 
\label{firstpage} 
 
\begin{abstract} 
We present spectroscopic observations from the {\it Spitzer Space Telescope} 
of six carbon-rich AGB stars in the Sagittarius Dwarf Spheroidal Galaxy (Sgr 
dSph) and two foreground Galactic carbon stars.  The band strengths of the 
observed C$_2$H$_2$ and SiC features are very similar to those observed in 
Galactic AGB stars.  The metallicities are estimated from an empirical relation
between the acetylene optical depth and the strength of the SiC feature. The
metallicities are higher than those of the LMC, and close to Galactic
values.  While the high metallicity could imply an age of around 1\,Gyr, for
the dusty AGB stars, the pulsation periods suggest ages in excess of 2 or 3
Gyr.  We fit the spectra of the observed stars using the DUSTY radiative
transfer model and determine their dust mass-loss rates to be in the range
1.0--3.3$\times 10^{-8} $M$_{\odot}$yr$^{-1}$.  The two Galactic foreground
carbon-rich AGB stars are located at the far side of the solar circle, beyond
the Galactic Centre.  One of these two stars show the strongest SiC feature
in our present Local Group sample. 
\end{abstract}

 

\section{Introduction} 
The Asymptotic Giant Branch (AGB) phase occurs during the late stages of the 
evolution of low- and intermediate-mass stars.  This phase is characterised by 
intense mass loss and leads to the formation of a circumstellar envelope made 
of gas and dust.  The molecular composition of this envelope is dependent on 
the C/O abundance ratio. The CO molecule is very stable and unreactive, and if 
$\rm C/O> 1$, all the oxygen is trapped in CO.  The envelope (and star) is then 
``carbon-rich''; amorphous carbon dominates the dust, while (after H$_2$) CO 
and C$_2$H$_2$ dominate its gas.  The``oxygen-rich'' AGB stars are 
characterised by silicate dust and molecules such as SiO, OH, H$_2$O. Third 
dredge-up brings the carbon produced by the triple-$\alpha$ reactions to the 
surface, increasing the C/O ratio and over time can change a star from 
oxygen-rich to carbon-rich. 
 
The mass loss from AGB stars is one of the main agents for the chemical 
evolution of galaxies.  It expels the products of nuclear reactions 
in the core of the star into the interstellar medium.  The mass 
loss from AGB stars contributes roughly half of all the gas recycled by 
stars (Maeder 1992), and up to nearly 90\%\ of the dust (Gehrz 1989).  Mass 
loss from AGB stars is also one of the main sources of carbon in the universe, 
together with Wolf-Rayet stars and supernovae (Dray et al.\ 2003).

The mechanisms driving this mass-loss process are not fully understood. It
is thought to be a two-step process: pulsations from the star lead to the
formation of dust, and then radiation pressure accelerates the dust grains to
the escape velocity. The gas is then also expelled due to friction with the
dust grains. The effect of metallicity on the mass-loss rates from AGB stars
has been discussed at some length. Some works proposed that
 mass-loss rates should be lower in metal-poor environments, as less dust is
expected to form, so that radiation pressure would be less efficient in
driving the mass loss (Bowen \& Willson 1991; Zijlstra 2004).  This
hypothesis on the metallicity effect on mass loss has recently been tested
using spectroscopy from \textit{Spitzer} (e.g.\ Sloan et al.\ 2006, 2008;
Zijlstra et al.\ 2006b; Lagadec et al.\ 2007; Matsuura et al.\ 2007; hereafter
``SZLM'' will refer to all five of these papers).  Against early 
expectations, the dust mass-loss
rates of carbon stars in metal-poor galaxies from the Local Group appear to
be similar to the ones measured in the Galaxy. In contrast, oxygen-rich
stars in metal-poor environments do appear to have lower dust mass-loss
rates. This suggests that carbon is important in triggering the superwind
(Lagadec \& Zijlstra 2008).  Theoretical models confirm that the mass-loss
rates of carbon stars should not depend on metallicity (Wachter et al.
2008). Mattson et al.\ (2008) argue that the pulsation energy of the star can
drive a strong wind at low metallicity, but the dependence on chemistry
suggests this is not the dominant effect within the observed metallicity
range.
 
The dusty circumstellar envelope surrounding AGB stars absorbs the light from 
the central star and re-emits it in the infrared.  Furthermore, spectral 
signatures of dust and molecules appear in this wavelength range. Spectroscopic 
observations of AGB stars in the thermal infrared are therefore vital for 
the study of the dusty envelopes around these stars.  Atmospheric 
absorption prevents ground-based mid-infrared observations outside two 
windows around 10 and 20$\mu$m.  The \textit{Spitzer Space Telescope} (Werner 
et al.\ 2004), with its high sensitivity and mid-infrared wavelength coverage, 
has proven to be a valuable tool for the study of dusty envelopes around 
extragalactic AGB stars. 
 
To study the mass loss from evolved stars at low metallicity, we are 
undertaking a Spitzer spectroscopic survey of mass-losing AGB stars in 
different Local Group galaxies. We have already presented results for the 
Magellanic Clouds and Fornax (SZLM; Matsuura et al.\ 2006; Groenewegen et 
al.\ 2007).  Here we present \textit{Spitzer} spectra of eight AGB stars in 
the direction of the Sagittarius Dwarf Spheroidal (Sgr dSph) 
galaxy.\footnote{This galaxy is also known as the Sagittarius Dwarf Elliptical
Galaxy (SagDEG), but it should not be confused with the Sagittarius Dwarf
Irregular Galaxy (SagDIG).} This study aims at studying the circumstellar 
properties of these AGB stars. 
 
\section{Target selection} 
We selected eight AGB candidates in the direction of the Sgr dSph based on 
their near-infrared colours.  These stars were not spectroscopically confirmed 
carbon stars before this study. The core of the Sgr dSph has a distance 
modulus estimated to be 17.02$\pm$0.19 (Mateo et al.\ 1995) and is located 
behind the Galactic Bulge.  It is being disrupted by the Galaxy so that its 
tidal tail surrounds the Milky Way.  It contains several stellar populations. 
The dominant one has a metallicity in the range [Fe/H]=$-$0.4 to $-$0.7 and 
an age of 8.0$\pm$1.5 Gyr (Bellazzini et al.\ 2006).  A second population is 
younger and more metal-rich with [Fe/H]=$-$0.25 for the most metal-rich 
objects (Zijlstra et al.\ 2006a).  These populations span the range of 
metallicities observed in the  Magellanic Clouds.

The eight selected targets are a subsample of the stars presented by Lagadec 
et al.\ (2008).  Table\,\ref{targets.dat} lists some characteristics of these 
stars.  For convenience, we will use the names published previously. The stars 
were selected to span a wide range in $J-K$ colour, i.e.\ a wide range of 
optical depth due to dust in their envelopes. 
 
\begin{figure} 
\begin{center} 
\includegraphics[width=9cm]{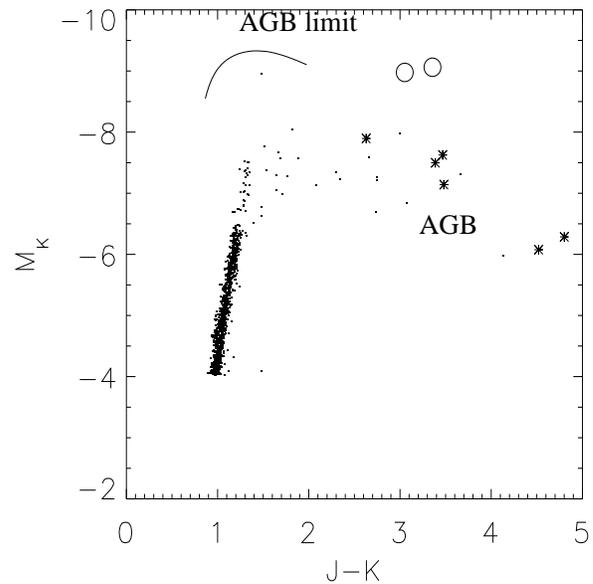} 
\caption{\label{mk_jk}The $M_K$ vs $J-K$ diagram of the observed sample. 
Crosses  represent the observed Sgr dSph targets and the large circle the 
foreground targets. Points represent the stellar population of Sgr dSph. 
$M_K$ is the absolute magnitude and was calculated assuming a distance 
modulus estimated to be 17.02$\pm$0.19 (Mateo et al.\ 1995).} 
\end{center} 
\end{figure}

Fig.\ \ref{mk_jk} displays the eight observed stars in an $M_K$ vs $J-K$ 
diagram.  Asterisks represent the observed stars in the Sgr dSph. To show that 
the observed stars are all AGB stars, we overplot the distribution of Sgr 
field stars.  Lagadec \& Zijlstra (2008) explain the selection of these field 
stars.  To verify that the observed stars belong to Sgr dSph, we performed 
radial velocity measurements: these indicate that six out of the eight observed 
stars belong to Sgr dSph but the remaining two are foreground stars (see 
Sec.\ \ref{rad_vel}).

\section{Observations and data reduction} 
%
\label{obs} 
\subsection{Spitzer observations} 
The observations were made with the InfraRed Spectrograph (IRS, Houck et al.\ 
2004), on board the \textit{Spitzer Space Telescope}. We used the Short-Low 
(SL) and Long-Low (LL) modules to cover the wavelength range 5-38$\mu$m.  The 
SL and LL modules are each divided in two spectral segments, together known as 
SL2, SL1 , LL2 and LL1; a ``bonus'' order covering the overlap between the two 
modules is also available.  The data reduction is similar to that described by 
Zijlstra et al.\ (2006b).  The raw spectra were processed through the {\it 
Spitzer} pipeline S15. We replaced the bad pixels by values estimated from 
neighbouring pixels. The sky was subtracted by differencing images 
aperture by aperture in SL and nod by nod in LL. We used the software tools
available in SPICE (the {\it Spitzer} IRS Custom Extractor) to extract the 
spectra. The flux calibration made use of the reference stars HR 6348 
(K0 III) in SL and HR 6348, HD 166780 (K4 III) and HD 173511 (K5 III) in LL. 
The spectra were individually extracted from the individual images. Both nods 
in both apertures were then joined simultaneously, recalculating the errors 
in the process by comparing the nods.  The different nods were averaged, using 
the differences to estimate the errors.  The different spectral segments were 
combined using scalar multiplication to eliminate the discontinuities due to 
flux lost because of pointing errors.  The different segments were also 
trimmed to remove dubious data at their edges.  We also retained the data in
the bonus order where it was valid.  These steps resulted in a standard 
wavelength calibration accuracy of 0.06\,$\mu$m in SL and 0.15\,$\mu$m in LL. 
 
\begin{figure*} 
\includegraphics[width=18cm]{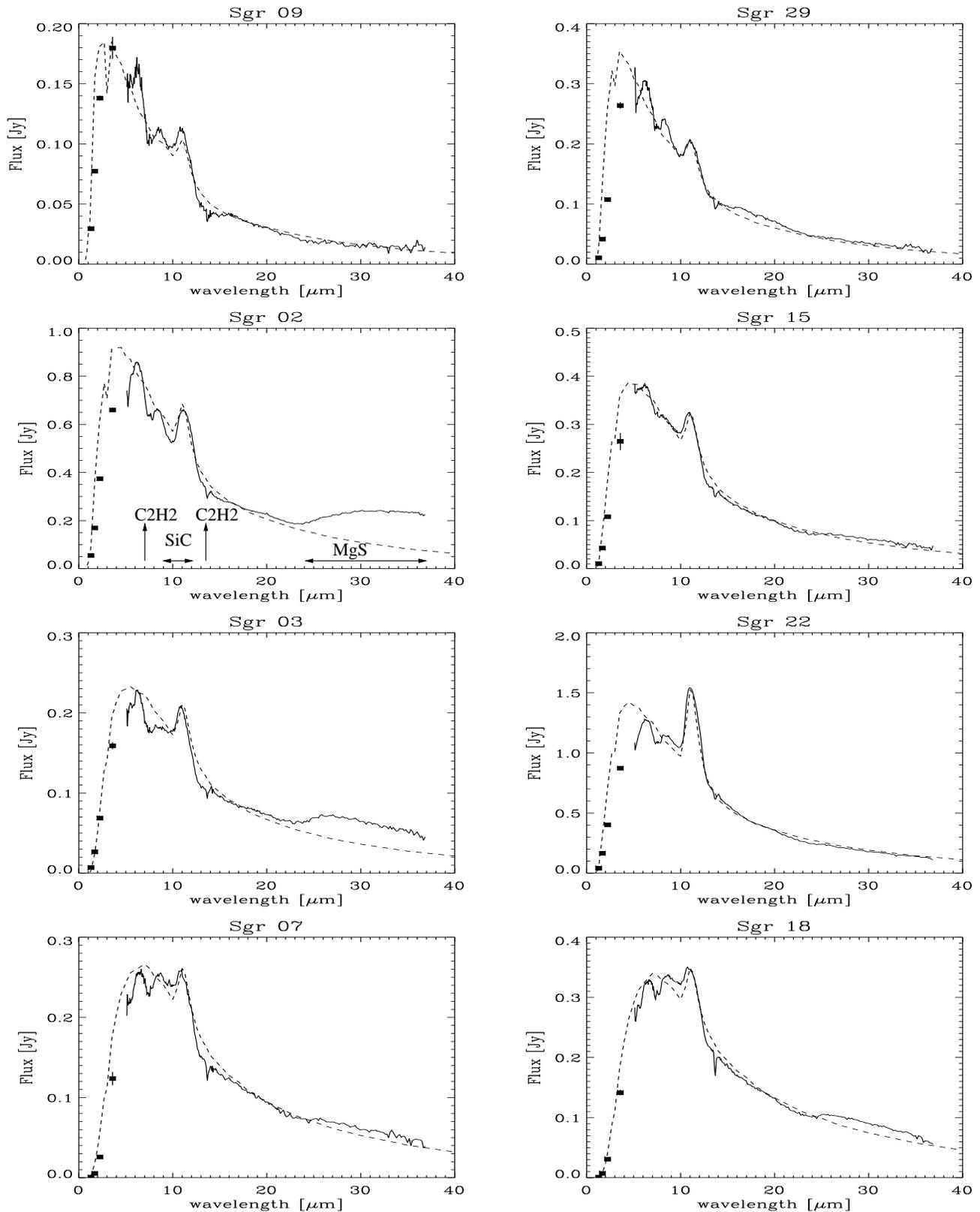} 
\caption{\label{spectra_sgr} Spitzer/IRS spectra of the eight observed carbon  
stars, ordered by dust temperature. The main dust and gas features are 
labelled on the spectrum of Sgr 02.  The dashed lines represent the SED fits 
obtained with DUSTY (Sec.\ \ref{dmlr})} 
\end{figure*}

 
Fig.\ \ref{spectra_sgr} presents the spectra of the observed AGB stars, 
ordered  by their [6.4]-[9.3] colours (see Sec.\ 5).  The molecular bands and dust 
features discussed below identify all eight objects as carbon-rich stars.

\subsection{Near-infrared photometry} 
Observations in the near-infrared are important to obtain a reliable estimate 
of the luminosity.  Due to the variability, these observations are best 
obtained close to the same epoch as the spectra.  Multi-epoch $JHKL$ 
photometry was obtained at the Australian National University (ANU) 2.3-m 
telescope at Siding Spring Observatory
(SSO) in Australia.  The filters used were centred at 1.28$\mu$m ($J$), 
1.68$\mu$m ($H$), 2.22$\mu$m ($K$) and 3.59$\mu$m ($L$).  Groenewegen et al.\
(2007) describe the observations.  Table\,\ref{targets.dat} presents the 
measured $JHKL$ magnitudes at the epoch of the \textit{Spitzer} observations. 
 
\begin{figure*} 
\includegraphics[width=12.5cm]{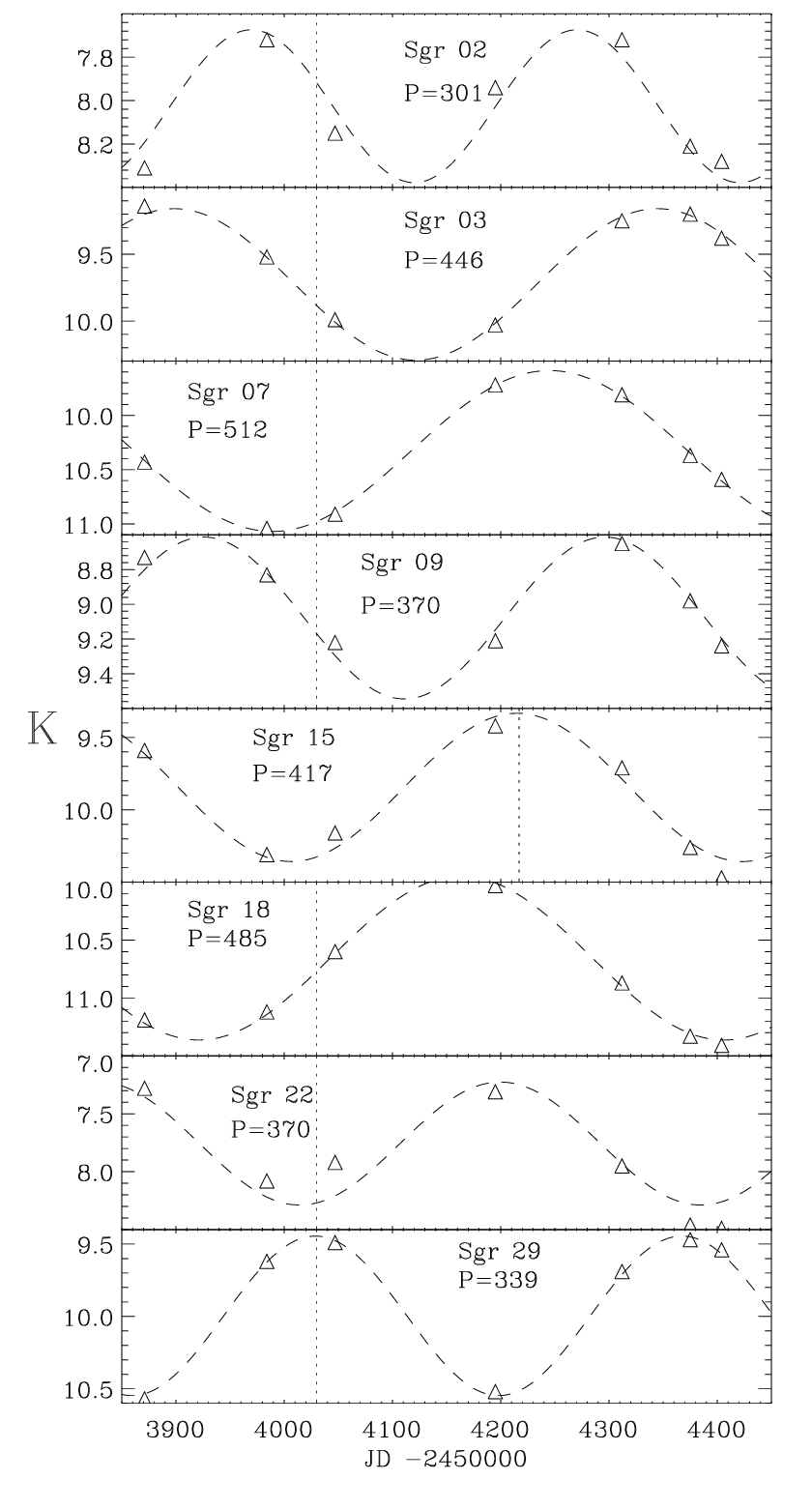} 
\caption{\label{lightcurves} Light curves of the eight observed carbon stars.  
These were obtained using a sinusoidal fit to the $K$-band magnitudes.  The 
vertical lines show the dates of our \textit{Spitzer} observations.} 
\end{figure*}

The multi-epoch observations, taken before the \textit{Spitzer} observations, 
were used to study the near-infrared variability of the sources.  Light curves 
were obtained by fitting a sine wave to the $K$-band data.  
Fig.\,\ref{lightcurves} displays these light curves. 
  

\subsection{Radial velocities} 
\label{rad_vel} 
Radial velocities were determined for six of the eight carbon stars using 
optical spectra obtained with the Dual Beam Spectrograph on the 2.3-m
SSO telescope.  The spectra have a resolution of 0.48 $\rm \AA$/pixel for 
four objects and 3.7 $\rm \AA$/pixel for the remaining two objects (see 
Table\,\ref{helradvel}).  The higher resolution spectra were cross-correlated 
with the local carbon star X Vel for which a spectrum at a resolution of 
0.5 $\rm \AA$/pixel was also obtained.  A radial velocity of 
$-5.4\,\rm km\,s^{-1}$ was adopted for X Vel (determined from high-resolution 
echelle spectra which were cross correlated against the radial velocity 
standard $\alpha$ Cet with a heliocentric radial velocity of 
$-$25.8\,km\,s$^{-1}$).  Cross-correlation of the lower-resolution spectra of 
the remaining two stars was used to obtain their radial velocities. 
Table\,\ref{helradvel} gives the final heliocentric radial velocities and 
their errors (as given by the IRAF task FXCOR).  
 
\begin{figure} 
\includegraphics[width=8cm,clip=true]{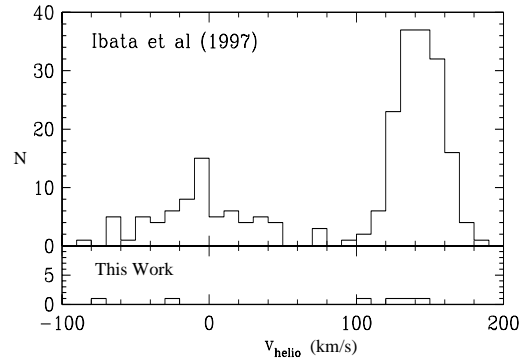} 
\caption{\label{radvel}Top panel:  Radial velocity distribution of stars in 
the  direction of the Sgr dSph galaxy (from Ibata et al.\ 1997).  Bottom 
panel: radial velocity distribution of six of our targets.  Two of these
stars, Sgr\,02 and Sgr\,22, are clearly foreground members of the Galaxy.}
\end{figure}

Fig.\ \ref{radvel} presents a histogram of the radial velocities of carbon 
stars in the direction of Sgr dSph from Ibata et al.\ (1997).  There are 
obviously two groups of stars:  those with heliocentric radial velocities of 
100 $<$ v$_{\rm helio}$(km/s) $<$ 190 which belong to the Sgr dSph; and those 
with v$_{\rm helio}$(km/s) $<$ 50 which belong to the Milky Way Galaxy.  Two
stars in our sample, Sgr\,02 and Sgr\,22, clearly belong to the Milky 
Way Galaxy, while the other stars are members of the Sgr dSph.  The 
properties of the two foreground stars are discussed in Sec.\ \ref{foreground}
 
We used other distance estimates to ascertain the Sgr dSph membership of 
the two stars without radial velocity measurements (see Sec.\ \ref{distance}).

\begin{table*} 
\caption[]{\label{targets.dat} Observed Sgr dSph targets: 
adopted names, coordinates, photometry and distance. 
$J$, $H$, $K$ and $L$ are taken from near-simultaneous measurements at SSO. 
The periods were obtained by sine-fit to the $K$-band observations.  The 
distance was estimated using the method described in Sec.\ \ref{distance}.} 
\begin{center} 
\begin{tabular}{llllllllllllllll} 
\hline 
Adopted name &  2MASS name & IRAS name& RA & Dec   
& $J$ & $H$ & $K$ & L&P &phase &D\\ 
  &&  & \multicolumn{2}{c}{(J2000)}&mag  & mag & mag & mag &d& &kpc\\ 
\hline 
 
 Sgr\,02& 18414350 $-$3307166&18384$-$3310&18 41 43.50 & $-$33 07 16.6 & 11.095 &   9.421 &  8.043  & 6.501 & 301 &0.65&17.1  \\ 
 Sgr\,03& 18443095 $-$3037098&18413$-$3040&18 44 30.96 & $-$30 37 09.8 & 13.360 &  11.436 &  9.879  & 8.047 & 446 &0.36&31.6 \\ 
 Sgr\,07& 18465160 $-$2845489&18436$-$2849&18 46 51.60 & $-$28 45 48.9 & 15.464 &  13.121 & 10.944  & 8.320 & 512 &0.06&34.1 \\ 
 Sgr\,09& 18514105 $-$3003377&            &18 51 41.05 & $-$30 03 37.7 & 11.753 &  10.263 &  9.124  & 7.913& 370 &0.40&30.3  \\ 
 Sgr\,15& 18584385 $-$2956551&18555$-$3001&18 58 43.85 & $-$29 56 55.1 & 12.862 &  10.921 &  9.394  & 7.496 & 417 &1.00&28.0 \\ 
 Sgr\,18& 19043562 $-$3112564&19013$-$3117&19 04 35.62 & $-$31 12 56.4 & 15.535 &  12.753 & 10.734  & 8.171 & 485 &0.42&35.1 \\ 
 Sgr\,22& 19103987 $-$3228373&19074$-$3233&19 10 39.87 & $-$32 28 37.3 & 11.315 &   9.454 &  7.959  & 6.196 & 370 &0.02&16.0 \\ 
 Sgr\,29& 19485065 $-$3058317&            &19 48 50.65 & $-$30 58 31.9 & 12.909 &  11.014 &  9.522  & 7.838  & 339& 1.00&31.5 \\ 
 
\hline \\ 
\end{tabular} 
\end{center} 
\end{table*} 
 

\begin{table} 
\caption[]{\label{helradvel}Heliocentric radial velocities for six of the  
observed targets as measured from observations at SSO.} 
\begin{center} 
\begin{tabular}{lrrrllllllllllll} 
\hline 
 Target  & Resolution &wavelength range&  v$_{\rm helio}$ & v$_{\rm error}$\\ 
 &$\rm \AA$/pixel & $\AA$&km/s & km/s \\  
\hline 
 Sgr\,02   &0.48 &8100-8900&$-$22.8      &1.1 \\ 
 Sgr\,03   &3.7 &6800-9200 &128.1        &11.1\\ 
 Sgr\,09   &0.48 &8100-8900&144.9        &1.6 \\ 
 Sgr\,15   & 3.7 &6800-9200&136.4        &12.2 \\ 
 Sgr\,22   &0.48 &8100-8900&-73.1        &2.1 \\ 
 Sgr\,29   & 0.48&8100-8900&106.3        &2.5 \\ 
\hline \\ 
\end{tabular} 
\end{center} 
\end{table} 
  
\section{Description of the spectra} 
\label{spectra} 
\begin{figure} 
\includegraphics[width=8cm,clip=true]{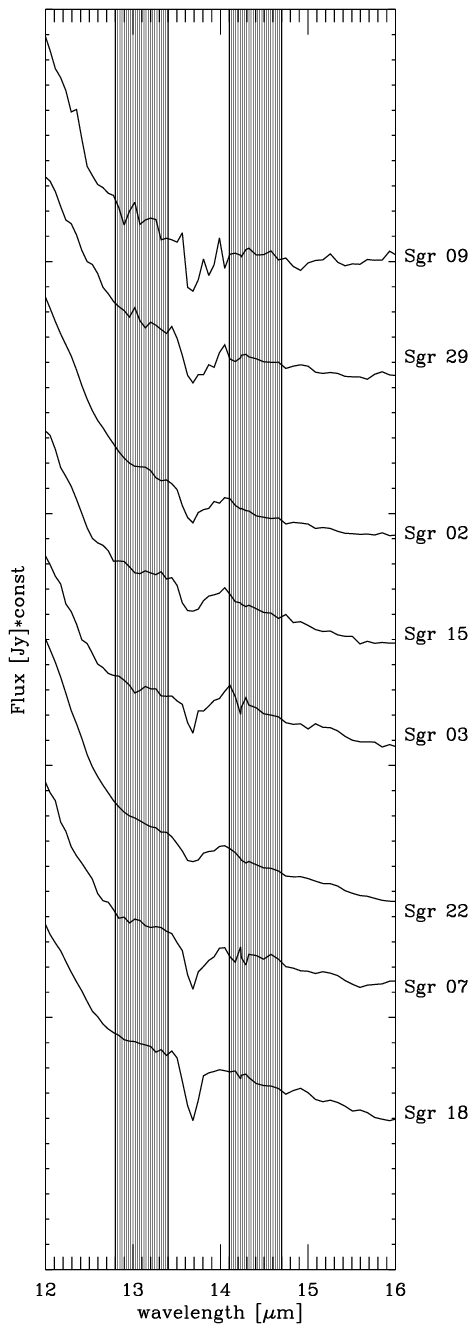} 
\caption{\label{c2h2} Spitzer spectra around 13.7$\mu$m of the observed stars, 
showing the C$_2$H$_2$ absorption.  The two grey bands represent the continuum 
used to measure the band strengths.} 
\end{figure}

The IRS spectra of the eight observed stars (Fig.\ \ref{spectra_sgr}) show 
both dust and molecular emission.  All the observed stars are carbon-rich and 
have spectra typical of AGB stars.  The spectra of all stars clearly show 
absorption features from C$_2$H$_2$ at 7.5 and 13.7\,$\mu$m (Fig.\ \ref{c2h2}). 
SiC dust produces the 11.3\,$\mu$m feature detected in all the spectra.  Some 
of them (Sgr\,02, Sgr\,03, Sgr\,07 and Sgr\,18) show a broad emission feature 
around 30\,$\mu$m attributed to MgS (Hony et al., 2002).  This feature might 
also be present in Sgr\,09 and Sgr\,22.  A drop is observed for all stars at 
the blue edge of the spectra.  It is due to several molecules, most notably
 C$_3$ and CO (Zijlstra et al.\ 2006b). 

Sgr\,22 displays a very strong SiC emission feature, stronger than any 
carbon-rich AGB star of which we are aware, including the Galactic sample
observed by the \textit{Infrared Space Observatory} and described by
Sloan et al.\ (2006) or the Local Group galaxies observed by \textit{Spitzer} 
(see upper panel of Fig.\ \ref{dust_6.eps}). 
 
A weak feature attributed to the stretching vibration of a carbonyl group 
(X-CO) has been observed in some LMC AGB stars (Zijlstra et al.\ 2006b) at 
5.8$\mu$m. This feature is present in the spectra of Sgr\,09 and Sgr\,18, and 
possibly also in Sgr\,02, Sgr\,03, Sgr\,07 and Sgr\,15. 

\section{Colours and band strengths}  
\label{col_strength} 
\begin{figure} 
\includegraphics[width=8cm,clip=true]{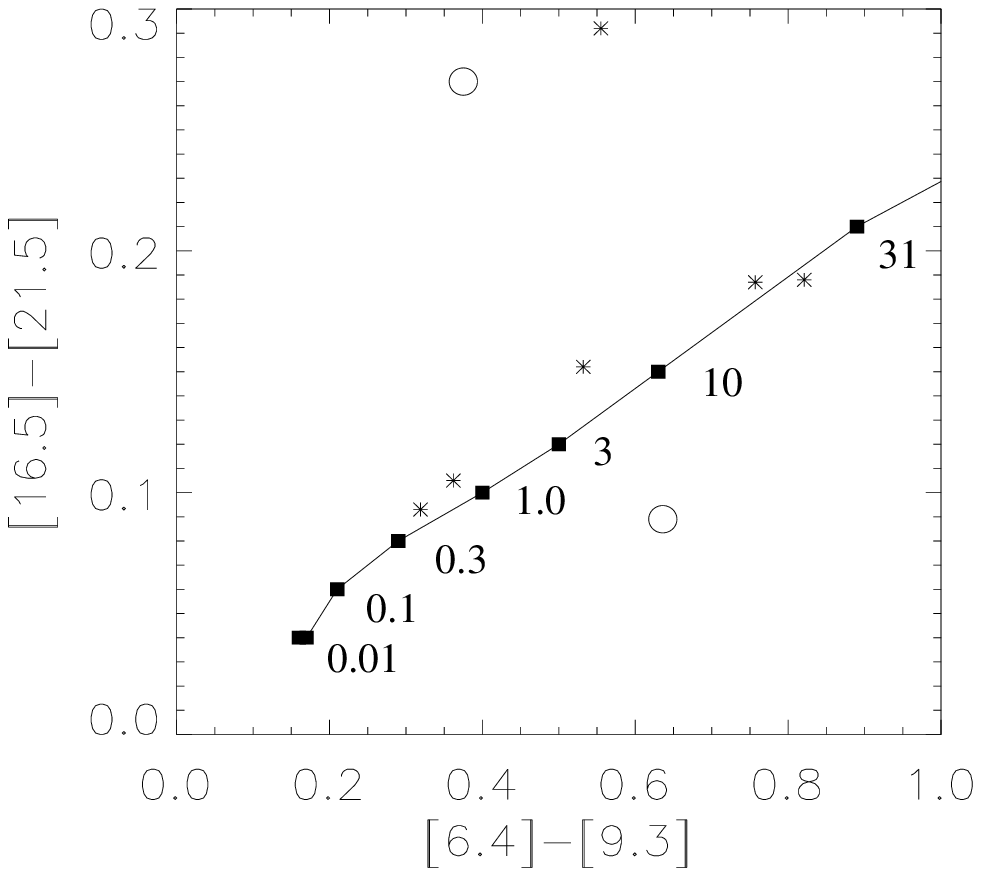} 
\caption{\label{col_man.eps} [16.5]$-$[21.5] vs. [6.4]$-$[9.3] colour 
diagram.  Asterisks and open circles represent the stars we observed.  
The solid line and filled squares represent a series of \textit{DUSTY} 
models, as described by Zijlstra et al.\ (2006a). 
The optical depth at 1$\mu$m is indicated for each model} 
\end{figure}

To determine colours of the observed stars, we applied the so-called 
``Manchester System'' (Sloan et al.\ 2006; Zijlstra et al.\ 2006b), using 
four narrow bands selected to represent the continuum at different 
wavelengths.  Using this method, we derive two colours.  Zijlstra 
et al.\ (2006b) showed that the [6.4]$-$[9.3] colour is a good estimate of 
the dust optical depth, while the [16.5]$-$[21.5] colour provides an 
estimate of the dust temperature.  The [6.4]$-$[9.3] colour shows a linear 
relation with the measured mass-loss rates (Groenewegen et al.\ 2007; 
Matsuura et al.\ 2007; Sloan et al.\ 2008).  Table  \ref{ew.dat} lists the 
blackbody temperature derived from this [16.5]$-$[21.5] colour.  To test that 
our choice of continuum colours is reasonable, we plotted [16.5]$-$[21.5] vs.\
[6.4]-[9.3] (Fig.\ \ref{col_man.eps}). For five of the observed stars, the 
continua are consistent.  But three stars (Sgr\,02, Sgr\,03 and Sgr\,22) are 
outliers.  Sgr\,02, Sgr\,03 are particularly red at [16.5]$-$[21.5].
 
For all the observed stars, we measured the strength of the main features 
(Fig.\ \ref{spectra_sgr}).  The continuum underlying each feature is defined 
using small wavelength ranges on the blue and red sides of the feature.  The 
continuum is then defined using a straight line between the blue and red 
continuum values.  Because the red edge of the MgS feature (Sec.\
\ref{spectra}) is outside of the IRS spectral range, we used a blackbody with  
the temperature derived from the [16.5]$-$[21.5] to extrapolate the continuum 
under this dust feature.  After the definition of the continuum we determined 
the strengths of the features in two different ways.  For the dust features, 
we simply measured the ratio between the integrated feature flux and the 
continuum (F/C).  For the gas features, seen in absorption, we measured the 
equivalent widths.  We also measured the central wavelength of the SiC feature 
defined as the wavelength at which the flux on the blue side of the feature 
equals the flux on the red side.  Table \ref{ew.dat} lists the measured 
strengths and central wavelengths of the observed features.

\begin{table*} 
\caption[]{\label{ew.dat}  Colours measured using 
four narrow carbon stars continuum bands, strength of the molecular and dust  
features, in terms of either the equivalent width in microns, or the integrated 
flux-to-continuum ratio, F/C (Sect. \ref{col_strength}). The last column gives  
the continuum (black-body) temperature, derived from the [16.5]$-$[21.5] colour 
listed in Table \ref{targets.dat}} 
\begin{center} 
\begin{tabular}{lcccccccclllllll} 
\hline 
target       &[6.4]$-$[9.3] & [16.5]$-$[21.5]&  EW (7.5 $\mu$m)  & EW (13.7 $\mu$m) & F/C(SiC) & $\lambda_c$ (SiC)  &F/C (MgS)& T( \\ 
             & (mag)       &  (mag)       &($\mu$m)               &($\mu$m) & &($\mu$m) && (K)\\
\hline 
Sgr\,02    &  0.375 $\pm$0.012 & 0.270 $\pm$0.010 & 0.148$\pm$   0.002 &  0.035 $\pm$ 0.005 &0.267$\pm$ 0.005& 11.27 $\pm$ 0.03 & 0.639 $\pm$ 0.012&   515.$\pm$  16. \\ 
Sgr\,03   &  0.555 $\pm$0.004 & 0.292 $\pm$0.009 & 0.115$\pm$   0.005 &  0.020 $\pm$ 0.008 &0.226$\pm$ 0.007& 11.07 $\pm$ 0.05 & 0.401 $\pm$ 0.012&   480.$\pm$  12. \\ 
Sgr\,07    &  0.757 $\pm$0.004 & 0.187 $\pm$0.015 & 0.078$\pm$   0.005 &  0.036 $\pm$ 0.002 &0.157$\pm$ 0.004& 11.17 $\pm$ 0.04 & 0.180 $\pm$ 0.018&   716.$\pm$  48. \\ 
Sgr\,09    &  0.319 $\pm$0.016 & 0.093 $\pm$0.018 & 0.237$\pm$   0.009 &  0.064 $\pm$ 0.011 &0.278$\pm$ 0.008& 11.14 $\pm$ 0.04 & 0.050 $\pm$ 0.026&  1385.$\pm$ 225. \\ 
Sgr\,15    &  0.532 $\pm$0.006 & 0.152 $\pm$0.014 & 0.061$\pm$   0.004 &  0.018 $\pm$ 0.005 &0.222$\pm$ 0.005& 11.11 $\pm$ 0.03 & 0.172 $\pm$ 0.018&   864.$\pm$  66. \\ 
Sgr\,18    &  0.821 $\pm$0.002 & 0.188 $\pm$0.013 & 0.068$\pm$   0.004 &  0.034 $\pm$ 0.003 &0.158$\pm$ 0.003& 11.18 $\pm$ 0.03 & 0.234 $\pm$ 0.015&   710.$\pm$  41. \\ 
Sgr\,22    &  0.636 $\pm$0.005 & 0.089 $\pm$0.016 & 0.093$\pm$   0.002 &  0.023 $\pm$ 0.002 &0.410$\pm$ 0.006& 11.20 $\pm$ 0.02 & 0.055 $\pm$ 0.020&  1457.$\pm$ 225. \\ 
Sgr\,29    &  0.362 $\pm$0.010 & 0.105 $\pm$0.017 & 0.131$\pm$   0.004 &  0.044 $\pm$ 0.010 &0.189$\pm$ 0.008& 11.22 $\pm$ 0.06 & 0.001 $\pm$ 0.021&  1230.$\pm$ 166. \\ 
                  
\hline \\ 
\end{tabular} 
\end{center} 
\end{table*} 


\section{Circumstellar properties} 

\subsection{Gas} 
As mentioned in Sec.\ \ref{spectra}, the main gas absorption features observed 
in our spectra at 7.5$\mu$m and 13.7$\mu$m are due to C$_2$H$_2$.  We have 
shown (SZLM, Matsuura et al.\ 2006) that in metal-poor environments, the 
C$_2$H$_2$ absorption becomes stronger.  This can be explained by the fact 
that at low metallicity, the initial oxygen abundance is smaller than for 
Galactic stars, and the dredge-up of carbon during the AGB leads to 
higher C/O ratios.  
 
\begin{figure} 
\includegraphics[width=8cm,clip=true]{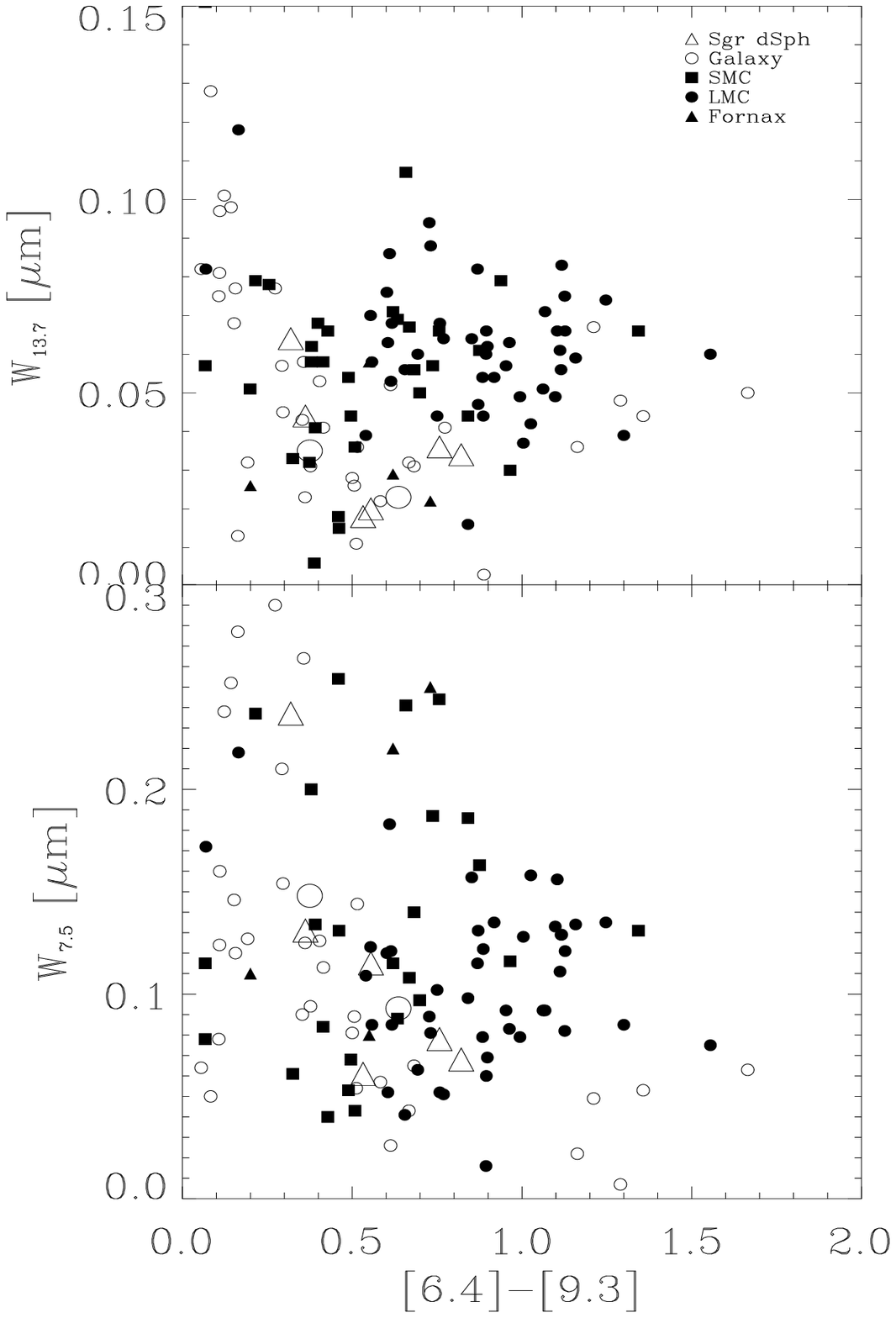} 
\caption{\label{gas_6.eps} The equivalent width of the C$_2$H$_2$ features at  
13.7$\mu$m (upper panel) and 7.5$\mu$m (lower panel) as a function of 
[6.4]$-$[9.3] colour.  The plot combines stars from the present sample and 
stars from previous Local Group observations: open triangles represent the 
current Sgr dSph sample, filled squares the SMC stars from Sloan et al.\ (2006) 
and Lagadec et al.\ (2007) samples, open circles the ISO Galactic sample 
defined by Sloan et al.\ (2006), filled circles the LMC sample of Zijlstra et 
al.\ (2006b) and Leisenring et al.\ (2008), and triangles stars from the 
Matsuura et al.\ (2007) Fornax sample.} 
\end{figure}

Fig.\ \ref{gas_6.eps} shows the equivalent width of these features as a 
function of the [6.4]$-$[9.3] colour for the current sample in comparison to 
the stars from other Local Group galaxies.  The plot relates the strength of 
the molecular band to the optical depth (or dust mass-loss rates) of the 
envelopes of the observed stars.  The C$_2$H$_2$ equivalent widths of the 
Sgr dSph sample are in the lower range of the observed strengths.  They are 
similar to or a little higher than the values found in Galactic stars of 
similar [6.4]$-$[9.3], but are generally weaker than found in the SMC, LMC 
and Fornax.  
 


\begin{figure} 
\includegraphics[width=8cm,clip=true]{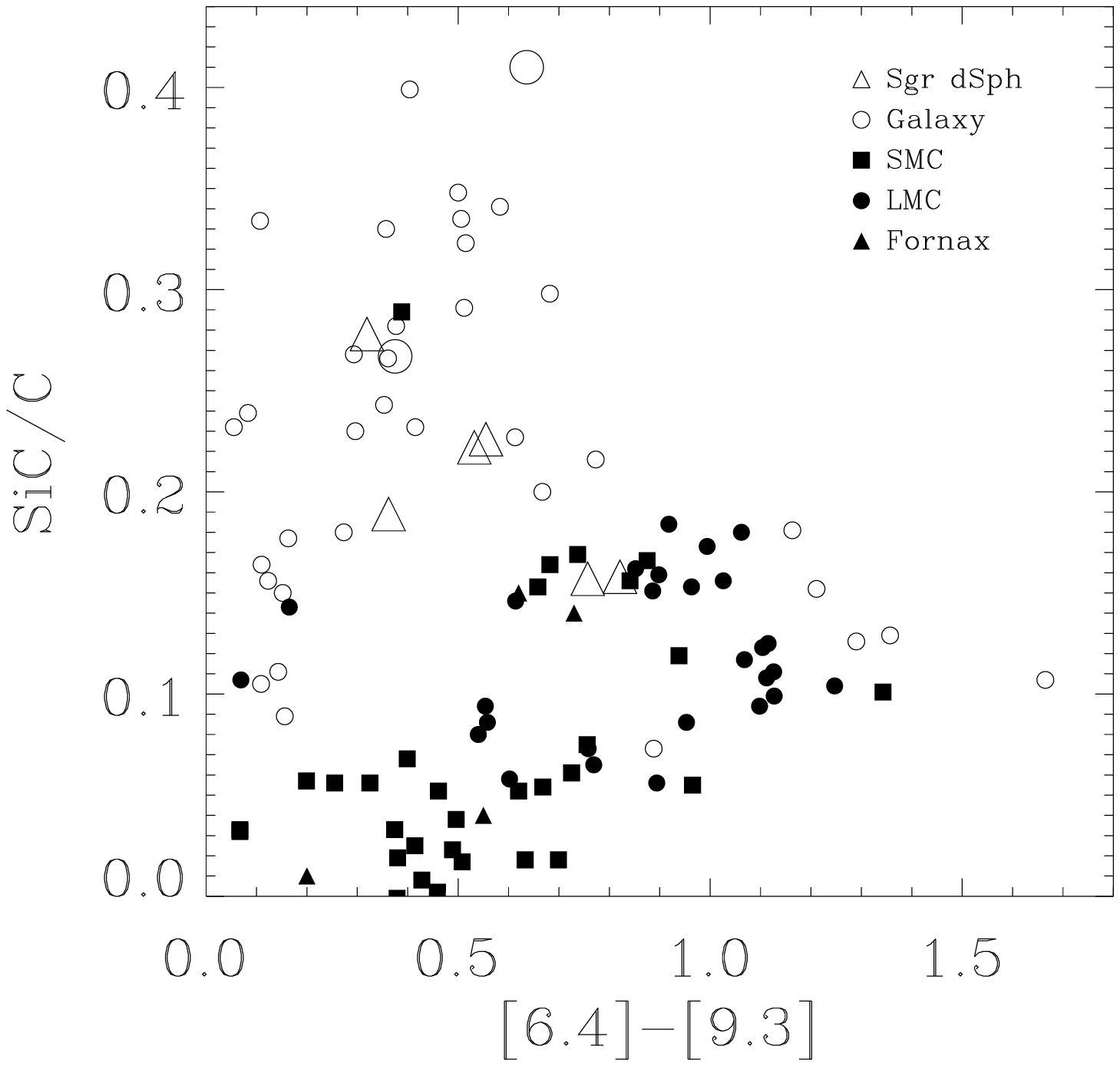} 
\caption{\label{dust_6.eps} The strength of the SiC features as a function
of the [6.4]$-$[9.3] colour.  L/C(SiC) is the integrated-flux-to-continuum  
ratio of the SiC feature (Sec.\ \ref{col_strength}).  Comparison data are 
from the literature are as in Fig.\ref{gas_6.eps}} 
\end{figure} 

\subsection{Dust} 
\label{dust} 

As mentioned in Sec.\ \ref{spectra}, the main dust features observed in the 
spectra are due to SiC and MgS.  The featureless continuum arises primarily
from emission from amorphous carbon emission.  Fig.\ \ref{dust_6.eps} shows 
the strength of the SiC and MgS features as a function of the [6.4]$-$[9.3] 
colour. 

The properties of the SiC feature have been extensively studied previously
(e.g.\ Speck et al.\ 2005, Leisenring et al.\ 2008).  Lagadec et al.\
(2007) have shown that the relationship between the strength of the SiC
feature and the optical depth of the envelope varies according to the
metallicity of the host galaxy, using a sample of stars in the SMC, LMC and
the Galaxy.  Fig.\ \ref{dust_6.eps} illustrates the trend of increasing SiC 
feature strength for the Galactic stars when [6.4]$-$[9.3] $\lsim$0.5 and a 
gradual decrease for redder stars.  Such a trend is observed for carbon stars 
in the LMC, but the inflexion occurs for redder colours, around $\sim$1. Very 
few SMC and Fornax stars with [6.4]$-$[9.3]\,$>1$ have been observed, but such 
stars in metal-poor galaxies appear to follow the same trend as LMC stars and 
have much weaker SiC features at a given optical depth.  One SMC star, GM 780, 
is an outlier with a very strong SiC emission feature. The stars from the Sgr
dSph are located at similar positions as the Galactic ones, but with a SiC 
feature slightly weaker than those of Galactic stars with the same [6.4]-[9.3]
colour.  As Si is not produced in AGB stars, this slight difference appears
to indicate that the Si abundance in the observed Sgr dSph stars is between 
the Si abundance of Fornax/SMC/LMC and the Galaxy.

 
 
\begin{figure} 

\includegraphics[width=8cm,clip=true]{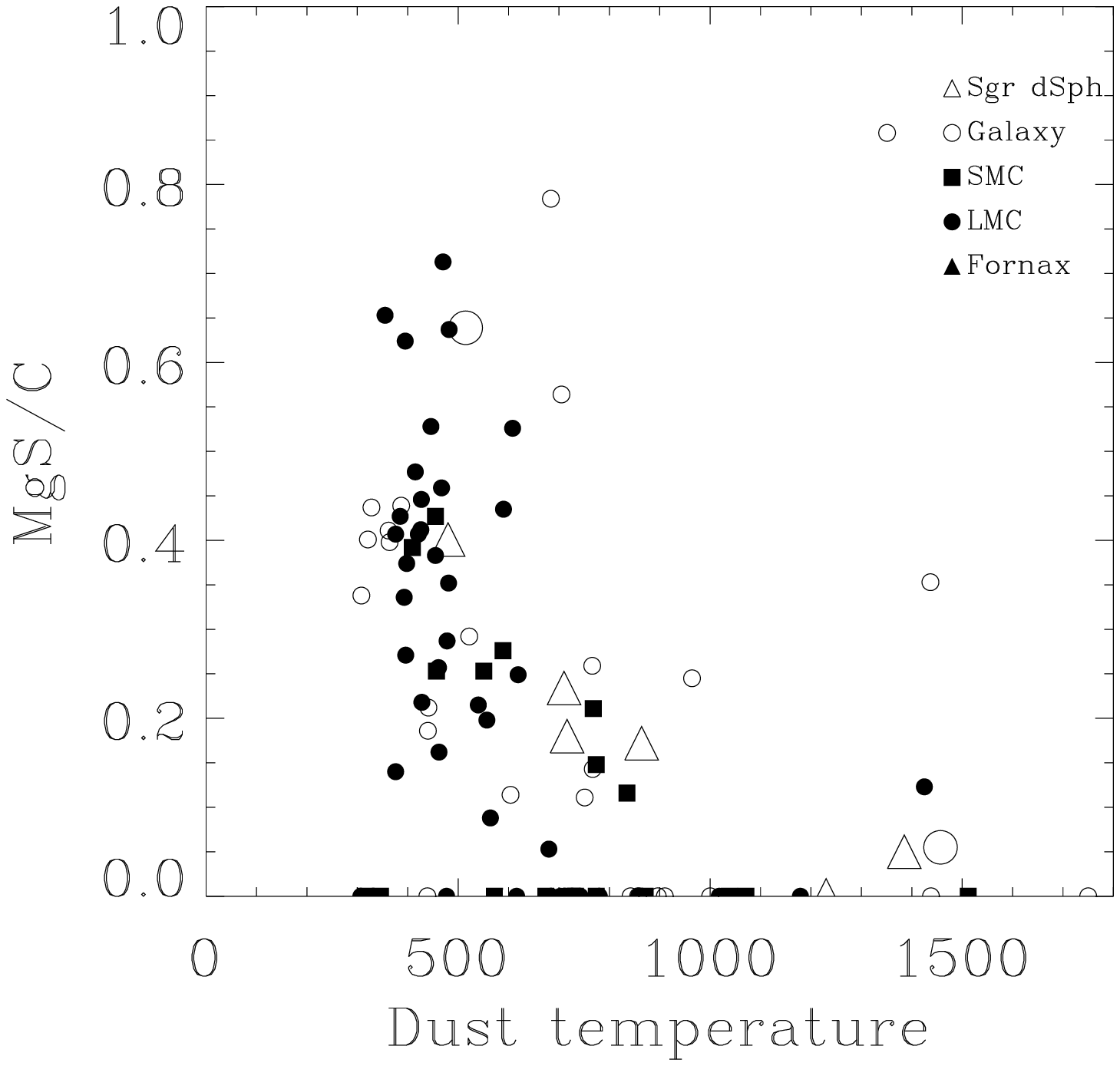} 
\caption{\label{mgs_dustT.ps} Strength of the MgS feature as a function of 
dust temperature estimated from the [16.5]$-$[21.5] colour.  Literature data 
as in Fig.\ref{gas_6.eps}} 
\end{figure}

 
Fig.\ \ref{mgs_dustT.ps} shows the strength of the MgS feature as a function 
of the dust temperature.  As already observed for AGB stars in other Local 
Group galaxies (Sloan et al.\ 2006, Zijlstra et al.\ 2006b, Lagadec et al.\ 
2007), this feature tends to appear mostly in the envelopes of stars with cool 
dust.  In general, the formation process of MgS begins around 600\,K and is 
complete around 300\,K.   

\section{Distance estimates} 
\label{distance} 
To check whether the observed stars are members of Sgr dSph, we estimated their
distances using two methods.  The distance to Sgr dSph being well established, 
our sample allows us to test different distance estimation methods.  

The first one uses the infrared colours of the observed stars, following the 
method described by Sloan et al.\ (2008).  They have shown, from a sample of 
carbon stars in the Magellanic Clouds, that: 
 
\begin{equation} 
M_K=-9.18+0.395(J-K) 
\end{equation}  
 
and 
 
\begin{equation} 
M_{9.3\mu m}=\sum a_i([6.4]-[9.3])^i 
\end{equation}  
 
where $a_0=-8.81$, $a_1=-6.56$ and $a_2=2.74$.  Using these relations and the 
infrared colours of the observed carbon stars, we can obtain two estimates of 
their distances.          

The second method makes use of the period-luminosity relation for carbon 
Miras as described by Feast et al.\ (2006): 
 
\begin{equation} 
M_{\rm bol}=-2.54 \times log P+2.06 
\end{equation} 

We converted all of the photometry to the SAAO system and dereddened the 
near-infrared data as described by Lagadec et al.\ (2008).  The reddening 
was estimated using the extinction maps by Schlegel et al.\ (1998).  For 
the observed stars, A$_J$$\sim$0.1 and A$_K$$\sim$0.04.  Using standard 
conversion factors, we estimated that A$_{6\mu m}$$<$0.01 and that the 
reddening was lower than  a percent at longer wavelengths.  We thus did 
not deredden our spectra.  We then derived the bolometric magnitudes using 
the equation for bolometric correction derived by Whitelock et al.\ (2006):
 
\begin{eqnarray} 
\nonumber {\rm BC_K} & = & +\, 0.972 + 2.9292\times(J-K) 
  -1.1144\times(J-K)^2 \\ 
 & & +0.1595\times(J-K)^3 -9.5689\,103(J-K)^4 
\end{eqnarray} 
\noindent  

Table \ref{distances} displays the mean values of the estimated distance 
for each star using the dereddened infrared colours and the 
period-luminosity relation. This confirms that the stars Sgr\,07 and Sgr\,18
belong to Sgr dSph.
 
\begin{table} 
\caption[]{\label{distances}Distances estimated from the infrared colours and 
the period-luminosity relation.} 
\begin{center} 
\begin{tabular}{lcccccccclllllll} 
\hline 
target       &D$_{IR}$ (kpc)& D$_{PL}$(kpc) \\ 
 
\hline 
Sgr\,02    &17.1&12.9 \\ 
Sgr\,03    &31.6&33.2 \\ 
Sgr\,07    &34.1&41.4 \\ 
Sgr\,09    &30.3&25.2 \\ 
Sgr\,15    &28.0&25.8 \\ 
Sgr\,18    &35.1&32.7 \\ 
Sgr\,22    &16.0&12.9 \\ 
Sgr\,29    &31.5&25.1 \\ 
                  
\hline \\ 
\end{tabular} 
\end{center} 
\end{table} 

Mateo et al.\ (1995) derived a distance for the central region of Sgr dSph of
25.35$\pm$2.06 kpc.  A comparison of our two distance estimates indicates
that the best agreement with the distance to Sgr dSph is obtained with the
period-luminosity relation. Using the infrared colours gives distances to
the observed stars significantly higher than the well-determined distance to
the galaxy.  The distance estimated using the period-luminosity relation
gives results in perfect agreement for Sgr\,09, Sgr\,15 and Sgr\,29.  For the
three other stars, this method gives larger distances.  Variability will have 
some effect on the derived distances, but given that two of the three stars 
with large apparent distances are significantly redder in $J-K$ than any of 
the others it would seem quite possible that they are undergoing obscuration 
events of the type described by Whitelock et al.\ (2006) for Galactic Miras 
and which are also seen in at least one of the AGB stars in Fornax 
(Whitelock et al.\ 2009).  Photometric monitoring over long time scales of 
these stars would be needed to confirm that hypothesis, as stars with such 
behaviours show variations over periods of a few years.
 
\section{Dust mass-loss rates} 
\label{dmlr}
 
 
\begin{table*} 
\begin{center} 
\caption[]{\label{per_ml} Dust mass-loss rates for the observed stars and  
parameters obtained from our DUSTY models.  $\dot{M}_{\rm col}$ and 
$\dot{M}_{\rm dusty}$ are the dust mass-loss rates  (in M$_{\odot}$yr$^{-1}$) 
from Lagadec et al.\ 2008 and our DUSTY models, respectively.} 
 
\begin{center} 
\begin{tabular}{cccccccccccccccccccrlllllll} 
\hline 
Target& Luminosity &$\dot{M}_{\rm DUSTY}$ &T$_{eff}$& T$_{\rm in}$& SiC/AMC& 
      $\tau$ (0.55$\mu$m)& $\dot{M}_{\rm col}$ &V$_{\rm exp}$ \\ 
      & (L$_\odot)$   &  (10$^{-8}$M$_{\odot}$yr$^{-1}$) &(K)&(K)&&& 
              (10$^{-8}$M$_{\odot}$yr$^{-1}$)& km/s   \\ 
\hline 
Sgr\,02            &  11469 &  2.59          & 2800 &  1200  & 0.1   & 4.67   &     0.86  & 28         \\ 
Sgr\,03            &  4529  &  1.29          & 2800 &  1200  & 0.1   & 9.56   &     0.84  & 18         \\ 
Sgr\,07            &  4483  &  2.52          & 2800 &  1000  & 0.08  & 9.64   &     0.97  & 14         \\ 
Sgr\,09            &  6652  &  1.02          & 2800 &  1200  & 0.1   & 2.22   &     0.27  & 27         \\ 
Sgr\,15            &  8632  &  2.73          & 2800 &  1200  & 0.1   & 7.00   &     0.93  & 23         \\ 
Sgr\,18            &  5056  &  3.39          & 2800 &  1000  & 0.08  & 13.1   &     2.60  & 13         \\ 
Sgr\,22            &  13321 &  3.78          & 2800 &  1200  & 0.2   & 7.11   &     1.30  & 26         \\ 
Sgr\,29            &  9152  &  2.04          & 2800 &  1500  & 0.1   & 5.89   &     0.96  & 35         \\ 
 
\hline \\ 
\end{tabular} 
\end{center} 
\end{center} 
\end{table*} 
 
Lagadec et al.\ (2008) measured the dust mass-loss rates of the stars 
presented in this work.  These dust mass-loss rates ($\dot{M}_{\rm col}$), 
measured from near and mid-infrared colours are presented in 
Table \ref{per_ml}. 
 
The \textit{Spitzer} spectra, together with simultaneous near-infrared
photometric measurements give us an opportunity to better model the dust
emission from these stars.  We modeled the spectral energy distributions 
(SEDs), defined by the $JHKL$ flux and the \textit{Spitzer} spectra,
using the radiative transfer code DUSTY (Ivezi\'c \& Elitzur 1997).  Our
primary objective is to fit the dust continuum in order to estimate the 
optical depths and the mass-loss rates.  DUSTY solves the 1-D problem of 
radiation transport in a dusty environment.  For all our models, we assume 
that the irradiation comes from a point source (the central star) at the 
centre of s spherical dusty envelope.  The circumstellar envelope is filled 
with material from a radiatively driven wind.  All the stars are carbon-rich 
and display SiC in emission.  We thus assumed that the dust composition of 
the envelope is a mixture of amorphous carbon and SiC.  Optical properties 
for these dust grains are taken the work of Hanner (1988) and Pegouri\'e 
(1988) for amorphous carbon and SiC, respectively.  The grain-size 
distribution is assumed to be a typical MRN distribution, with a grain 
size $a$ varying from 0.0005 to 0.25$\mu$m distributed according to a power 
law with n($a$)$\propto$$a^{-q}$ with $q$=3.5 (Mathis et al.\ 1977).  The 
outer radius of the dust shell was set to 10$^3$ times the inner radius; 
this parameter has a negligible effect on our models.
 
To model the emission from the central star, we used a hydrostatic model 
including molecular opacities (Loidl et al.\ 2001, Groenewegen et al.\ 2007).
We used those hydrostatic models as an illuminating source only, we did not
attempt to model the molecular absorption features.  The C/O ratio is assumed 
to be 1.1 for the hydrostatic models, which is a typical value for Galactic 
AGB stars.  This ratio is not well known in the observed stars:  the C/O 
ratio of AGB stars appears to be higher in metal-poor environments than in 
the Galaxy (Matsuura et al.\ 2002, 2005).  But varying the C/O ratio mainly 
affects the depth of the molecular features (such as CO and C$_2$H$_2$). 
The radiative transfer model fits the dust emission, and the C/O ratio has 
little effect on the model.  We also performed some model calculations 
assuming that the central star emits as a blackbody, but these did not 
provide satisfactory fit to the near-infrared data, due to the absence of
molecular absorption.  The free parameters of the models are the dust 
temperature at the inner radius, the mass ratio of SiC to amorphous carbon 
dust and the central star effective temperature, T$_{\rm eff}$.   As one 
output, DUSTY provides an estimate of the terminal outflow velocity and the 
mass-loss rate for a luminosity of 10$^{4}$ L$_{\odot}$.  As the luminosity 
can be determined by scaling the emerging spectrum, we can then recalibrate 
the mass-loss rates accordingly, assuming a distance of 25 kpc for the six 
stars in Sgr dSph, and 17.1 and 16.0 kpc for Sgr\,02 and Sgr\,22 respectively. 
 
DUSTY gives total (gas+dust) mass-loss rates assuming a gas-to-dust ratio of 
200, but the dust is the dominant constraint on the fitting to the SED.
Therefore, we have divided the resulting mass-loss rates by the gas-to-dust
ratio and report only the dust mass-loss rate in Table \ref{per_ml}.  This
assumes that the gas and dust expansion velocities are the same.  Note that we 
had to make a number of assumptions in our models.  For example, we assume 
that the envelope  and the dust grains are spherical.
This cannot be the case in reality, so our fits cannot be perfect.  However,
the models achieve our primary aim, estimating the dust mass-loss rates.
These values are of the same order of magnitude as those derived by 
Lagadec et al.\ (2008).  The values found by Lagadec et al.\ (2008) are 
generally a factor of two or three smaller than the present ones.



For all the stars, the best fit is obtained with $T_{\rm eff}$=2800K.  The best 
model for most of the stars was obtained using a temperature at the inner 
radius of 1200\,K and a mixture of 10\% of SiC and 90\% of amorphous carbon.  
For two stars, we needed T$_{\rm in}$= 1000\,K to model the red part of the 
SED, and for another one we used T$_{\rm in}$= 1500\,K.  The very strong SiC 
feature observed in the spectrum of Sgr\,22 is very well fitted using 20\% 
of SiC. Sgr\,07 and Sgr\,18 have weaker SiC features, which can be fitted 
using a SiC mass fraction of 8\%. 

\begin{figure} 

\includegraphics[width=8cm,clip=true]{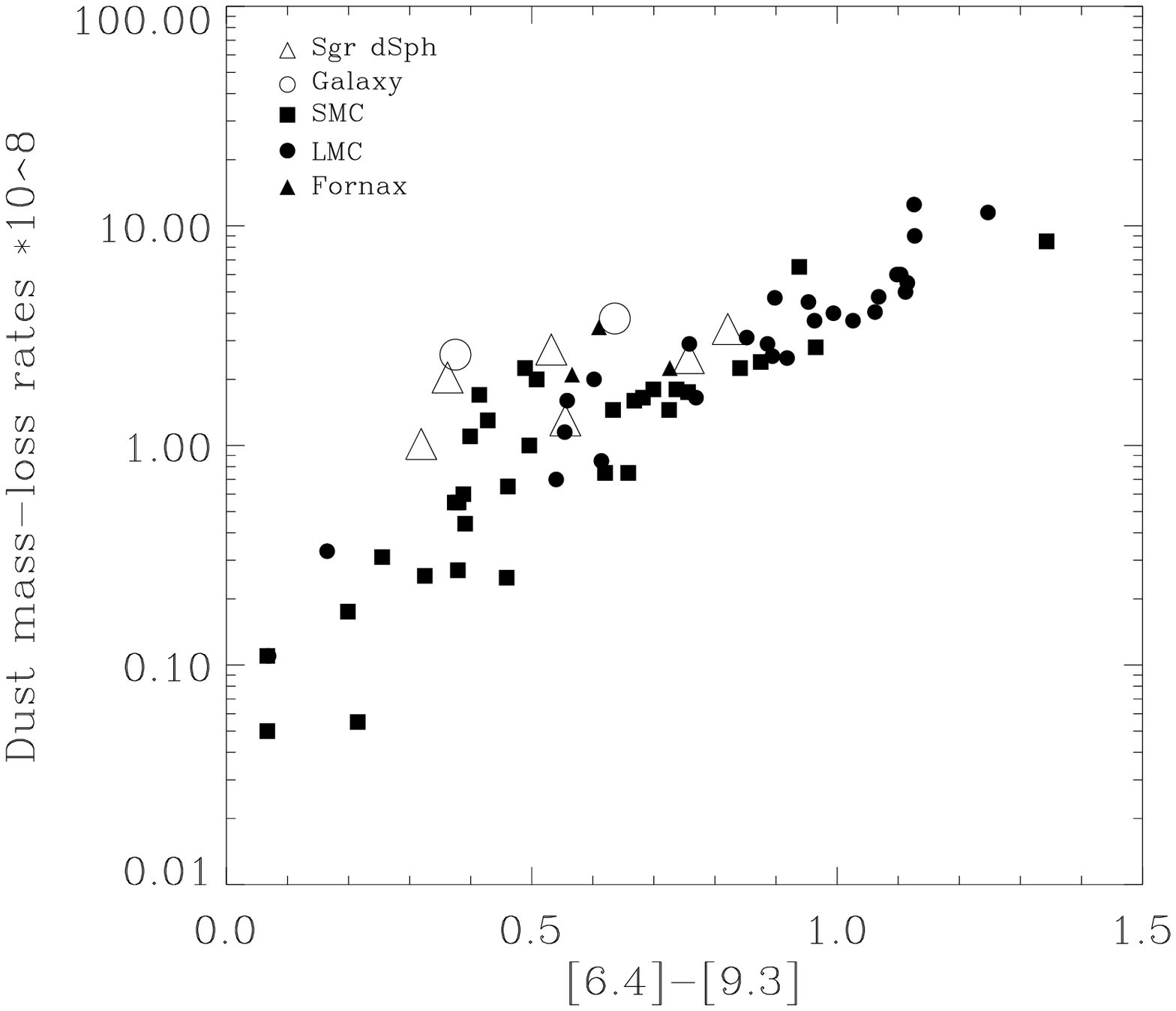} 
\caption{\label{mass_loss_col6.ps} Dust mass-loss rates as a function of  
the [6.4]$-$[9.3] colour.  The comparison data are described in 
Fig.\ \ref{gas_6.eps}} 
\end{figure}

 
We confirm the general correlation found by Groenewegen et al.\ (2007) and 
Matsuura et al.\ (2007) between the mass-loss rates and the [6.4]$-$[9.3] 
(Fig\,\ref{mass_loss_col6.ps}). But the stars we observed in the Galaxy and in 
Sgr dSph are on average offset towards higher mass-loss rates for the same 
colour, by about a factor of two. The two stars in Fornax show a (smaller) 
offset in the same direction.  All of these mass-loss rates of these stars were 
estimated from our DUSTY models, while the ones in the Magellanic Cloud stars 
were estimated using another radiative transfer model (Groenewegen et al., 
2007).  The models by Groenewegen et al.\ (2007) were made assuming a 
constant expansion velocity of 10 km s$^{-1}$ while DUSTY calculates an
expansion velocity for each model.  These calculated velocities are generally 
a factor of 2-3 higher than the ones assumed by Groenewegen et al., suggesting
that the difference in mass-loss rates arise from different assumptions made
by the models, rather than an intrinsic differences between the stars.  The
dust mass-loss rates determined by Lagadec et al.\ (2008) for the Sgr dSph 
stars nicely follow the relation with [6.4]$-$[9.3] colour found in previous
work.

\section{Discussion} 
\subsection{Estimating metallicities} 
\label{met_est} 
The survey of carbon-rich AGB stars in the Local Group has yielded 
\textit{Spitzer} IRS spectra of a large sample of stars in galaxies covering
a range of metallicities. 

The metallicity of extragalactic AGB stars is normally taken as that of the 
underlying dominant stellar population of the galaxy.  Individual 
metallicities of carbon stars are especially difficult to determine.
Generally, stellar metallicities are determined from the strengths of 
fine-structure lines from heavy elements in optical or 
near-infrared spectra.  The continua of AGB stars are very uncertain, due to 
the presence of multiple, overlapping molecular bands in their envelopes, 
e.g.\ CO and CN.  To model this continuum, a hydrostatic stellar atmosphere 
model is required, with a known surface gravity,  C/O abundance ratio, and 
effective temperature.  Such models have been made to determine the 
metallicity of Local Group AGB stars, but are very  time-consuming, and 
have been applied to only a small number of stars (de Laverny et al.\ 2006).   

The SiC feature is expected to depend on metallicity, through the Si 
abundance.  It may be possible to use the strength of this feature in
our \textit{Spitzer} spectra to estimate the metallicity of the stars
in our sample.  Simultaneously, the C$_2$H$_2$ abundance depends on the 
amount of free carbon, [(C$-$O)/O] (Lagadec \&\ Zijlstra 2008), which is 
expected to increase (on average) with decreasing metallicity (e.g.\ 
Matsuura et al.\ 2005).  Thus the \textit{Spitzer} spectra provide two 
means of determining metallicity.

\begin{figure} 
\includegraphics[width=8cm,clip=true]{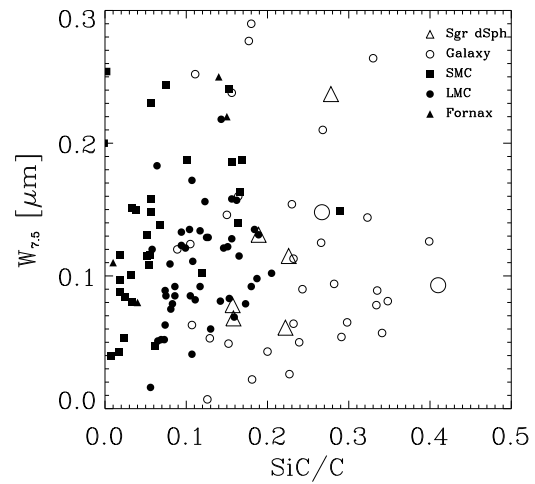} 
\caption{\label{sic_mgs.ps} Strength of the C$_2$H$_2$ 7.5$\mu$m feature as  
function  of the strength of the SiC feature.  The comparison data are as
described in Fig.\ref{gas_6.eps}} 
\end{figure}

Fig.\ \ref{sic_mgs.ps} shows the equivalent width of the C$_2$H$_2$ 
7.5\,$\mu$m feature versus the SiC flux-over-continuum ratio, for all the 
Local Group carbon stars we observed.  A clear separation is seen between the 
stars from different galaxies, i.e.\ different metallicities.  The Galactic 
stars, which are expected to be the most metal-rich, have the
largest SiC/C ratio for a given C$_2$H$_2$ 7.5\,$\mu$m equivalent width.  The 
LMC stars are spread out between the SMC and Galactic stars. The diagram 
shows a good (average) separation between the galaxies, in order of 
expected metallicity.  Stars from each individual galaxy show a significant 
spread,  which may be due to a spread in metallicity within a galaxy, or 
due to an evolutionary spread, as the stars will show an increasing carbon 
dredge-up over time.  But for the full carbon-star population, this diagram 
appears to provide a metallicity indicator. 
 
We can apply this diagram to the stars observed in this paper.  The two 
galactic foreground stars fall within the general population of Galactic 
stars, with one star (Sgr\,22) showing an unusually high SiC/C ratio which 
can be interpreted as arising from a high metallicity.  Surprisingly, the 
stars in the Sgr dSph are found between the Galactic and LMC stars, with 
an indicated metallicity higher than that of the LMC, although still 
sub-solar. Based on this diagram, the observed Sgr dSph carbon stars are 
the most metal-rich population after the Galactic stars.

\subsection{Metal-rich carbon stars in the Sgr dSph} 
 
The dominant giant branch population in the Sgr dSph is believed to have a 
metallicity ([Fe/H]) $\sim$$-0.55$ or less (Whitelock et al.\ 1996, 
Dudziak et al.\ 2000, Carraro et al.\ 2007) which is lower than that of the 
LMC.  The observed high metallicity implied for the dusty carbon stars in 
our Sgr dSph sample is therefore unexpected. 

Evidence for a more metal-rich stellar population in the Sgr dSph was first 
detected by Bonifacio et al.\ (2004), who, based on spectroscopy of ten giants, 
found the dominant population to have [Fe/H]\,$=-0.25$, much higher than what 
is found from typical AGB stars.  Zijlstra et al.\ (2006a) and Kniazev 
et al.\ (2008) show that one of the four planetary nebulae in the galaxy has 
the same high metallicity, confirming the existence of this population even 
if it does not dominate.  Chou et al.\ (2007) find a median metallicity in 
the core of the Sgr dSph of [Fe/H]$\,=-0.4$, similar to the LMC and again a 
little higher than typical AGB stars.  

The dusty carbon stars studied in this paper appear to be comparatively
metal-rich.  It is therefore natural to suggest that they could be an AGB 
counterpart of the population whose age has been estimated at 1\,Gyr 
(Bonifacio et al.\ 2004), the formation of which may have been triggered 
by the passage of the dwarf galaxy through the Galactic disc. However, 
their pulsation periods, which range from 310 to 512 days, imply 
considerably older ages.  For example there are three carbon Miras which 
have periods around 490 days (Nishida et al.\ 2000) in Magellanic Cloud 
Clusters which have ages about 1.6\,Gyr (Mucciarelli et al.\ 2007a,b; 
Glatt et al.\ 2008), whilst van Loon et al.\ (2003) suggest that the 
cluster KMHK 1603, which contains a carbon-rich Mira with a period of 
680 days, has an age of 0.9-1.0 Gyr.  The kinematics of Galactic 
carbon-rich Miras (Feast et al.\ 2006) is also consistent with increasing 
periods going with decreasing age.  The periods of the Sgr dSph Miras would 
therefore imply ages of 2 to 3 Gyr, or perhaps a little greater.  More work 
is required to reconcile these observations.

One can wonder why a relatively small galaxy would show the most metal-rich
stellar population in the Local Group after the two main spirals (e.g.\
Zijlstra et al.\ 2006a).  In general the dwarf spheroidals show a much 
steeper age-metallicity relation than do dwarf irregulars of the same 
luminosity.  Dwarf spheroidals are found around large galaxies while 
the dwarf irregulars are more isolated systems, showing the effect of
environment on enrichment history.  It is also of interest that of the 
satellites of the Milky Way, only the Sgr dSph has reached such high 
metallicity.  Fornax, otherwise a very similar system, contains carbon
stars with more metal-poor characteristics (Matsuura et al.\ 2007).  As 
the Sgr dSph is the nearest of the satellites (Fornax is around 130 kpc 
away), the effects of the Galaxy seem important.  
 
There are two possible explanations for the very high metallicities.  First, 
the stripping of the interstellar gas by a passage through the Milky Way may 
have allowed fast chemical evolution through stellar mass loss and supernovae. 
Second, the system may have captured gas from the Milky Way. The second option 
may have some support from the fact that the high metallicity is the same as 
that of the Galactic disk at the orbital distance of the Sgr dSph. 
 
\subsection{Two foreground carbon stars} 
\label{foreground} 
 

As discussed in Sec.\ \ref{rad_vel}, two stars in our sample, Sgr\,02 and 
Sgr\,22, belong to the Milky Way Galaxy while the other stars are members of 
the Sgr dSph.  These stars have been selected through their $J$ and $K$ 
colours and were not spectroscopically confirmed carbon stars before these 
observations.  The \textit{Spitzer} spectra clearly show that these stars are 
carbon-rich AGB stars.  Their distance is estimated to be 17.1 kpc and 
16.0 kpc for Sgr\,02 and Sgr\,22, respectively, from their infrared colours 
(see Sec.\ \ref{distance}).  Sgr\,22 displays the strongest SiC feature among 
all the stars from our present Local Group sample (including the Galaxy).  
The other foreground star, Sgr\,02, has a weaker SiC feature than Sgr\,22, but 
Fig.\ \ref{dust_6.eps} indicates that its SiC feature strength is similar to 
the ones observed in the Galaxy rather than in more metal-poor galaxies. 
 
The two Galactic stars are on line of sight to the Sgr dSph, and they are 
located on the far side of the Galaxy, close to the solar circle.  There are 
very few carbon stars known in the inner Galaxy: carbon star distribution 
begins around the solar circle and extends outwards (Le Bertre et al.\ 2001).  
The two stars are therefore located around the inner edge of the carbon star 
distribution and should be similar to carbon stars in the Solar neighbourhood.
However, their height above the Galactic plane is 3.6 and 4.9 kpc respectively. 
This is well beyond what is expected from the disc stars.  There is no clear 
explanation, and it would be very interesting to establish the metallicity of 
other halo carbon stars, some of which are known to be Mira variables 
(e.g.\ Totten et al.\ 2000; Mauron 2008).  The high metallicity suggests the 
stars escaped from the disc.  Whether the escape process was triggered by the 
Sgr dSph passage is not clear.

\section{Conclusions} 
 
We have presented a \textit{Spitzer} spectroscopic survey of six carbon-rich 
AGB stars in Sgr dSph and two serendipitously discovered Galactic carbon 
stars.  We used the low-resolution short-low and long-low mode of the IRS 
spectrometer to obtain mid-infrared spectra covering the range 5-38$\mu$m.  
These observations enabled us to study dust and gas features from the 
circumstellar envelopes around these stars.  We observed molecular absorption 
bands due to C$_2$H$_2$ at 7.5 and 13.7$\mu$m.  A weak absorption feature at 
14.3$\mu$m, due to HCN, is observed in two stars.  A weak absorption feature at 
5.8$\mu$m due to a carbonyl group (X-CO) is observed around two stars and 
might be present in the spectra of three others.  The continuum of all the 
observed stars is due to emission from amorphous carbon, which does not 
display any spectral features.  All the stars show a SiC feature at 
11.3$\mu$m.  A broad dust emission emission due to MgS is also observed for 
six of the stars around 30$\mu$m.

Radial velocity measurements show that two of our sample are actually carbon
stars.  The remaining six stars in our sample are members, as confirmed with
radial velocities, infrared colour-magnitude relations, and/or 
period-luminosity relations.  One of the two Galactic carbon stars displays
the strongest SiC feature ever observed.  Both Galactic stars are certainly
carbon stars located at the far side of the solar circle, but they have 
unexpectedly large distances from the Galactic plane.
 
We fitted the SEDs of all the observed spectra using the radiative 
transfer code DUSTY.  The estimated dust mass-loss rates are found to be in 
the range 1.0-3.3$\times 10^{-8}$M$_{\odot}$yr$^{-1}$. 
 
The observed strengths of C$_2$H$_2$ and SiC are very similar to the ones 
observed for Galactic carbon stars.  Stronger C$_2$H$_2$ and weaker SiC 
feature were expected in this metal-poor dwarf galaxy.  We have shown 
that the strength of these features depends on the metallicity and that
the observed stars have metallicities close to Galactic values.  This 
result is unexpected, as the usual measurements of the metallicity ([Fe/H]) of
the Sgr dSph range from $-0.4$ to $-0.7$.  The enhanced metallicity of the
observed carbon stars indicates that the interstellar medium in the Sgr dSph 
has been strongly enriched.

\section*{Acknowledgments} 
EL acknowledges support from a STFC rolling grant. EL thanks Xander Tielens 
and Kevin Volk for very helpful discussions during the preparation of this 
paper.  These observations were made with the {\it Spitzer Space
Telescope}, which is operated by JPL, California Institute of Technology 
under NASA contract 1407.  This research has made use of the SIMBAD and 
VIZIER databases, operated at the Centre de Donn\'{e}es astronomiques de 
Strasbourg, and the Infrared Science Archive at the Infrared Processing and 
Analysis Center, which is operated by JPL.

\label{lastpage} 
 
\end{document}